\documentclass[aps,prl,twocolumn,floatfix,showpacs]{revtex4-1}
\usepackage{amssymb}

\usepackage{epsfig}
\usepackage{amsmath}

\begin{document}

\title{Using the de Haas-van Alphen effect to map out the closed\\three-dimensional Fermi surface of natural graphite}
\author{J. M. \surname{Schneider}}
\author{B. A. \surname{Piot}}
\author{I. \surname{Sheikin}}
\author{D. K. \surname{Maude}}

\affiliation{Laboratoire National des Champs Magn\'etiques Intenses, CNRS-UJF-UPS-INSA, 38042 Grenoble, France}

\date{\today}

\begin{abstract}
The Fermi surface of graphite has been mapped out using de Haas van Alphen (dHvA) measurements at low temperature with
in-situ rotation. For tilt angles $\theta>60^{\circ}$ between the magnetic field and the c-axis, the majority electron
and hole dHvA periods no longer follow the $\cos(\theta)$ behavior demonstrating that graphite has a 3 dimensional
closed Fermi surface. The Fermi surface of graphite is accurately described by highly elongated ellipsoids. A
comparison with the calculated Fermi surface suggests that the SWM trigonal warping parameter $\gamma_3$ is
significantly larger than previously thought.
\end{abstract}


\maketitle

Graphite consists of Bernal stacked graphene layers with a weak inter layer coupling which leads to an in-plane
dispersion which depends on the momentum in the direction perpendicular to the layers, $k_z$. Graphite is a semi metal
with the carriers occupying a small region along the $H-K-H$ edge of the hexagonal Brillouin zone. The Slonczewski,
Weiss, and McClure (SWM) Hamiltonian with its seven tight binding parameters $\gamma_0,..., \gamma_5, \Delta$, is based
on group theoretical considerations and provides a remarkably accurate description of the band structure of
graphite~\cite{Slonczewski58,McClure60}. In a magnetic field, when trigonal warping is included ($\gamma_3\neq0$)
levels with orbital quantum number $n$ couple to levels with orbital quantum number $n+3$ and the Hamiltonian has
infinite order. However, the infinite matrix can be truncated as the eigen values converge rapidly~\cite{Nakao76}. The
validity of the SMW-model, has been extensively verified using many different experimental techniques \emph{e.g.}
Shubnikov-de Haas (SdH), de Haas-van Alphen (dHvA), thermopower, magneto-transmission, and magneto-reflectance
measurements
\cite{Soule58,Soule64,Williamson65,Schroeder68,Woollam70,Woollam71a,Orlita2008,Chuang09,Schneider2009,Zhu2009,Schneider2010a,Hubbard2011,Ubrig2011}.
However, recently claims~\cite{Luk04,Luk06} for the observation, in electrical transport measurements, of massless
two-dimensional (2D) charge carriers with a Dirac-like energy spectrum have caused much
controversy~\cite{Mikitik06,Lukyanchuk2010,Schneider2010}.

The Fermi surface of graphite has electron and hole majority carrier pockets with maximal extremal cross sections at
$k_z=0$ (electrons) $k_z \approx 0.35$ (holes). For both types of charge carriers the in-plane dispersion is parabolic
(massive fermions). Only at the H point ($k_z=0.5$) the in-plane dispersion is linear, similar to that of charge
carriers in graphene (massless Dirac fermions). At the H-point,there two possible extremal orbits. A minimal (neck)
orbit of the majority hole carriers, which gives rise to minority carrier effects ($\alpha$-surface) and a maximal
extremal orbit of the small ellipsoidal minority hole pocket ($\beta$ surface) which results from the intersection of
the two majority hole ellipsoids. The existence of the three minority carrier pockets at the K-point, the so called
outrigger pieces, was proposed by Nozi\`eres~\cite{Nozieres58}, however their existence is considered to be unlikely
due to the rather large value of $\gamma_3$, the SWM trigonal warping parameter, required.

In this Letter, we present a \emph{complete} map of the Fermi surface of natural graphite obtained from dHvA
measurements at low temperature ($T\approx 0.4$~K) with in-situ rotation. For tilt angles $\theta < 60^{\circ}$, the
dHvA periods of both the electrons and holes follow a $cos(\theta)$ dependence. While such a quasi-2D behavior is well
established in the literature, previous dHvA measurements ~\cite{Williamson65}, were unable to distinguish between a
highly elongated 3D ellipsoid and a cylindrical 2D Fermi surface. Our results at larger tilt angles demonstrate
unequivocally that graphite has a 3D closed Fermi surface which is accurately described by highly elongated ellipsoids
provided the spin splitting is included. A comparison of our data with the full SWM calculations allows us to refine
the SWM tight binding parameters, notably $\gamma_3$ is found to be significantly larger than previously thought.

\begin{figure}
\begin{center}
\includegraphics[width=7.5cm]{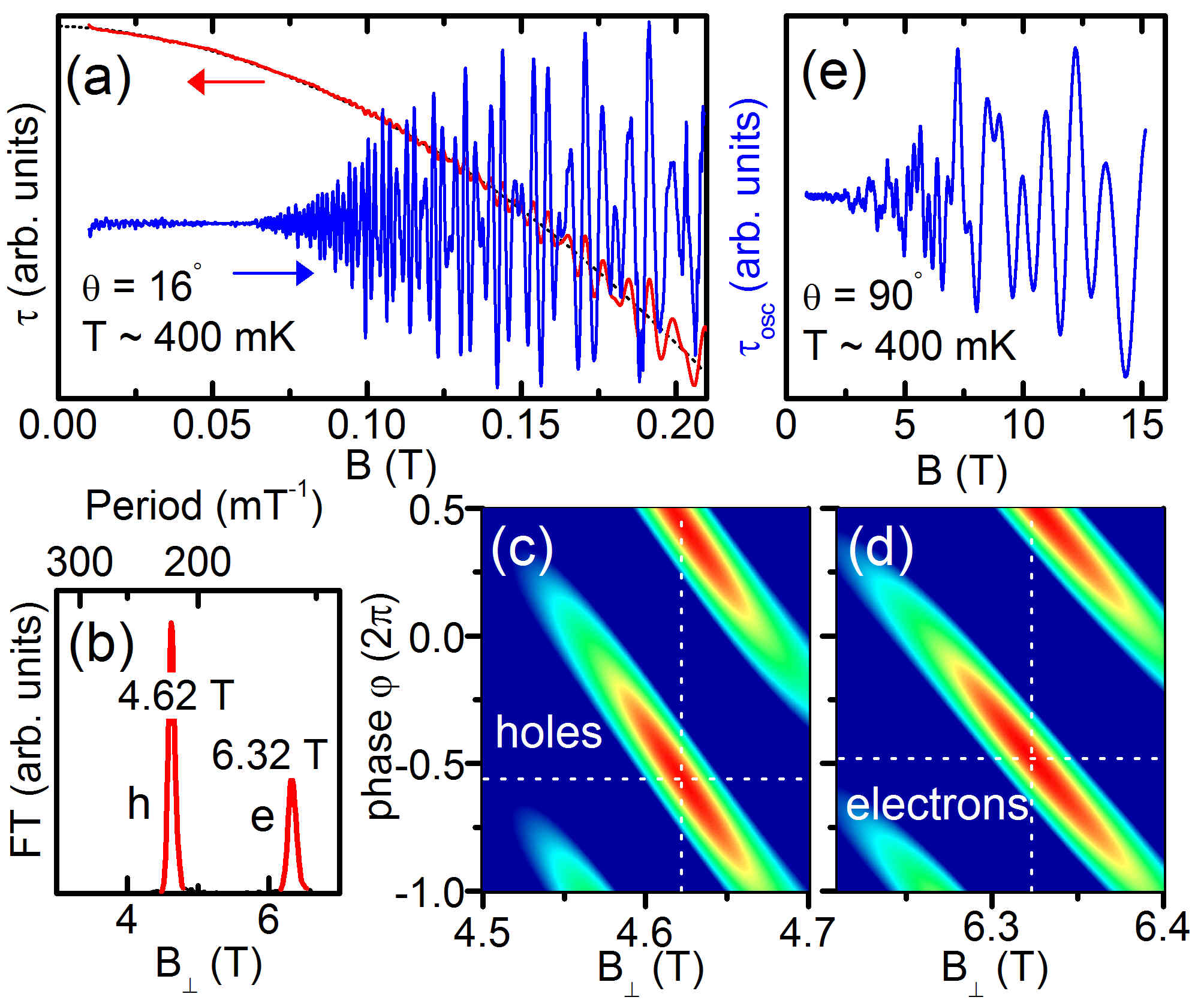}
\end{center}
\caption{(color online) (a) The torque $\tau$ and oscillatory torque $\tau_{osc}$ versus total magnetic field for
$\theta=16^{\circ}$. (b) Fourier transform of $\tau(1/B\cos(\theta))$. (c-d) The phase shift function
$\mathrm{\Re}\left[\mathrm{exp}\left(i\varphi\right) f(\mathrm{B_\perp})\right]$ as a function of the phase and
frequency. (e) Oscillatory torque $\tau_{osc}(B)$ for $B\parallel ab$ ($\theta = 90^{\circ}$).} \label{Fig1}
\end{figure}

For the dHvA measurements we used a mm-size piece of natural graphite, which was mounted on a CuBe cantilever which
forms the mobile plate of a capacitive torque meter. The capacitive torque signal was measured with a lock-in amplifier
using conventional phase sensitive detection at $5.3$~kHz. The measurements were performed using a $16$~T
superconducting magnet and a dilution fridge, equipped with an \emph{in situ} rotation stage. Fig.~\ref{Fig1} (a) shows
the torque $\tau$ as a function of the total magnetic field from $B=0-0.21$~T for a tilt angle $\theta = 16^{\circ}$.
The torque shows the expected dependence, $\tau(B) \propto -B^2$ (broken line), since $\tau = M B \sin(\theta)$ and the
magnetization $M$ depends linearly on the magnetic field. Superimposed on the large monotonic background, small quantum
oscillations are clearly visible, which reflect the oscillatory magnetization of the system as Landau levels pass
through the Fermi energy. The dHvA oscillations can be better observed in the oscillatory torque ($\tau_{osc}$). Here
the monotonic background has been removed by subtracting a smoothed (moving window average) data curve. The torque
$\tau\propto sin(\theta)$ so that we cannot directly access the quantum oscillations in perpendicular field. In order
to compare with our previous magnetotransport measurements we use the low angle $\theta = 16^{\circ}$ data writing
$B_\perp=B\cos(\theta)$. Fig.~\ref{Fig1}(a) shows $\tau_{osc}(B)$.

%
%
%
%
%

The phase and the frequency of the dHvA oscillations, were extracted from a Fourier analysis. The fundamental
frequencies $B_{F\perp}^{e}=6.32 \pm 0.1$~T and $B_{F\perp}^{h}=4.62 \pm 0.1$~T for electrons and holes respectively,
are obtained from the amplitude of the Fourier transform of $\tau_{osc}(1/B_\perp)$ (see Fig.~\ref{Fig1} (b)). In
Fig.~\ref{Fig1}(c) and (d) we plot the phase shift function
$K(\varphi,B_\perp)=\mathrm{\Re}\left[\mathrm{exp}\left(i\varphi\right)
f(\mathrm{B_\perp})\right]=\cos(\varphi-\varphi_0)F(B_\perp)$ as a function of the perpendicular magnetic field and the
phase. The fundamental frequency and the phase $\varphi_0$ can be found from the maxima in the $\varphi - B $ plane.
The phase values in units of $2\pi$ obtained are $\varphi_0^e=-(0.48\pm 0.1)$ and $\varphi_0^h=-(0.56\pm 0.1)$. The
phase $\varphi_0=\gamma-\delta$ with $\gamma=1/2$ for massive Fermions or $\gamma=0$ for massless Dirac fermions. For a
3D Fermi surface the curvature along $k_z$ gives $\delta=\pm 1/8$ for minimum/maximum extremal cross sections. In
contrast a cylindrical 2D Fermi surface gives $\delta=0$. We can therefore conclude, in agreement with recent dHvA
measurements on highly oriented pyrolytic graphite (HOPG)~\cite{Hubbard2011}, that both the electrons and holes are
massive Fermions with a parabolic energy spectrum (\emph{i.e.} $\gamma=1/2$). In Fig.~\ref{Fig1}(e) we show the
oscillatory torque $\tau_{osc} (B)$ in the $B \| ab$ configuration ($\theta=90^{\circ}$). The $B\parallel ab$
configuration can be found very precisely ($\delta \theta < 0.1^{\circ}$), since the magnetization background changes
sign at $\theta=90^{\circ}$. Well pronounced quantum oscillations are observed demonstrating unequivocally that the
Fermi surface of graphite is 3D and closed.

\begin{figure}
\begin{center}
\includegraphics[width= 7.5cm]{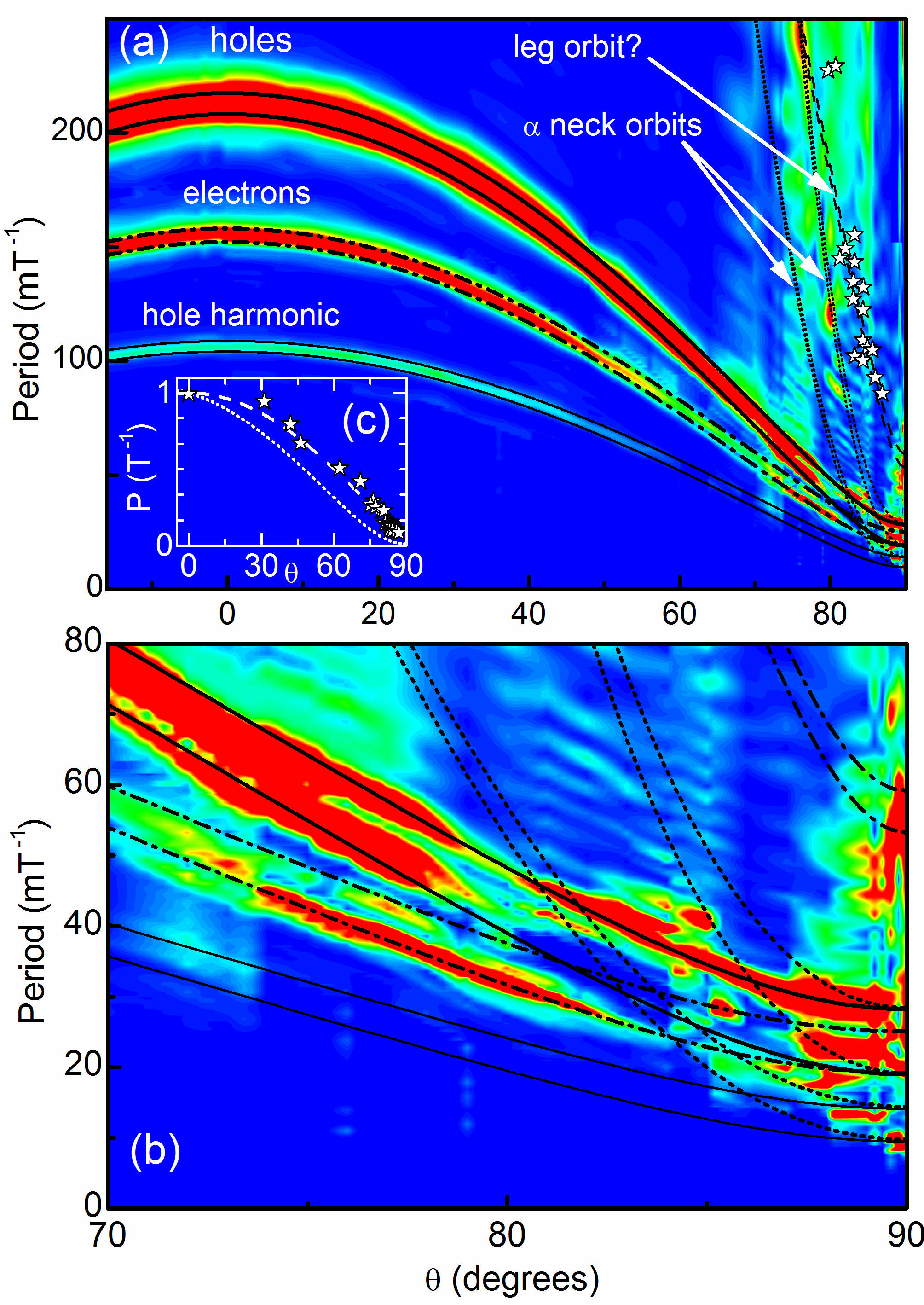}
\end{center}
\caption{(color online) (a-b) Color plot : Amplitude of the Fourier transform of $\tau_{osc} (1/B)$ as a function of
the period of the oscillations and the tilt angle ($\theta$). The calculated dHvA periods for ellipsoidal electron and
hole Fermi surfaces are also shown for hole (solid lines), hole harmonic (thin solid lines), and electrons (dot-dash).
The hole neck orbits with and without breakdown are shown as dotted lines. Woollam's minority carrier
data\cite{Woollam71b} is shown as symbols. The dashed line is the prediction for an ellipsoid. (c) Woollam's minority
carrier data plotted over an extended range. The prediction for an ellipsoid (dashed line) and for a neck orbit (dotted
line) are shown for comparison.}\label{Fig2}
\end{figure}

In order to map out the Fermi surface, we have performed systematic angle dependent measurements. In Fig.~\ref{Fig2}(a)
we plot the amplitude of the Fourier transform of $\tau_{osc} (1/B)$ as a function of the period of the oscillations)
and the tilt angle $\theta$. The hole and electron features, together with the hole harmonic, can clearly be
distinguished. For angles $\theta < 60^{\circ}$, the dHvA period for both electrons and holes follow the well
documented~\cite{Soule64,Williamson65} $\cos(\theta)$ dependence. Such a behavior is characteristic of either a 2D
material or very anisotropic material with an almost perfectly cylindrical Fermi surface. We can distinguish between
these two scenarios at higher tilt angle. The non cylindrical nature of the Fermi surface is clearly revealed for
$\theta \geq 60^{\circ}$ where deviations from the $\cos(\theta)$ behavior, are observed. Namely for $\theta \geq
75^{\circ}$ (see inset of Fig. \ref{Fig2}), the slope of the dHvA periods for both the electrons and holes features
changing dramatically reaching almost zero close to $\theta = 90^{\circ}$ (see Fig.~\ref{Fig2}(b)). In addition, the
hole feature clearly splits into two around $\theta=75^{\circ}$ due to a lifting of the spin degeneracy.

%
%
%
%

In a first approach the Fermi surface of graphite has been approximated using highly elongated ellipsoids. According to
the Lifshitz-Onsager relation \cite{onsager1952}, the fundamental frequencies $B_F=\hbar A / 2 \pi e$ are directly
proportional to the extremal cross sectional areas $A$ of the Fermi surface. For the maximal extremal orbits the area
is given by the intersection of a plane with the ellipse,
\begin{equation}\label{EqEllipsoid}
B_F \propto A=\pi ab/\sqrt{\sin ^{2}\theta + (a^2/b^2) \cos ^{2} \theta},
\end{equation}
where $a$ and $b$ are the semi-major and semi-minor axes of the ellipse and $\theta$ is the angle between the magnetic
field and the $c$-axis of the graphite crystal. For elongated ellipsoids ($a/b \geq 5$) this follows very closely $\pi
b^2/\cos \theta$ for $\theta<60^{\circ}$ \emph{i.e.} follows closely the behavior of a 2D cylindrical Fermi surface.

For our data, matters are further complicated by the observed spin splitting for $\theta > 70^{\circ}$. In order to
include spin splitting in our simple model, we note that the oscillatory term can be written as,
\begin{equation}\label{EqBfSpin}
\tau_{osc} \propto \cos\left(\frac{|E_f|}{\left(\frac{\hbar e}{m^*} \mp g^* \mu_B\right) B} + \phi\right) \equiv
\cos\left(\frac{B_{F}^{\uparrow\downarrow}}{B} + \phi\right),
\end{equation}
where $E_f$ is the Fermi energy, $m^*$ is the angle dependent effective mass, $g^*$ is the Land\'e g-factor, $\mu_B$
the Bohr magnetron and $\phi$ is a phase factor. From Equation~\ref{EqBfSpin} the frequency in the absence of spin
splitting is $B_F=|E_f| m^*/\hbar e$. Thus, we have a simple relation between the frequency (period) of oscillations
calculated for the ellipsoid and the expected frequencies when spin splitting is included,
\begin{equation}
1/B_{F}^{\uparrow\downarrow}=1/B_F \mp g^* \mu_B/2|E_f|
\end{equation}
so that the expected splitting of the period is simply $\pm g^* \mu_B/2|E_f|$ independent of the angle $\theta$. The
Fermi energy is $E_{F}^{h}\simeq-0.025$~eV for holes and $E_{F}^{e}=E_{f}^{h}-2\gamma_2\simeq 0.0246$~eV for electrons
with $\gamma_2=-0.0243$~eV \cite{Nakao76}. A reasonable fit to the observed splitting is obtained with $g^*=2.4$ for
electrons and $g^*=4.0$ for holes. The value for holes is considerably larger than the value of $g^*=2.5$ found for
both electrons and holes from magnetotransport~\cite{Schneider2010a}. It is not clear if this is due to the movement of
the Fermi energy which is not taken into account in our analysis, or if $g^*$ is really larger for holes. The cross
sectional area of the ellipsoids are obtained by fitting to the majority electron and hole frequencies at
$\theta=0^{\circ}$ and at high tilt angles $\theta>70^{\circ}$. The parameters used are summarized in Table~\ref{tab1}.
The results of such a fit are plotted in Fig.\ref{Fig2} as thick solid and dot-dash lines for the majority hole and
electron pockets. The simple model fits the experimental data remarkably well, reproducing the observed angular
dependence and the electron and hole spin splitting.

\begin{table}
\begin{center}
\begin{tabular}{ccccccc}
\hline\hline &$B_{f\perp}$~(T)&$A_{\perp}$&$A_{\parallel}$&$A_{\parallel}/A_{\perp}$&$A^{SWM}_{\perp}$\\
\hline
Maj. hole & 4.7 & 4.49 & 40.3 & 9.0 & 4.33\\
Maj. elec. & 6.45 & 6.15 & 43.1 & 7.0 & 6.26\\
Hole Leg(?)& $\simeq$1\footnote{Woollam\cite{Woollam71b}} & 0.95 & 17.0 & 17.8 & -\\
Hole neck& $\simeq$0.43\footnote{SWM calculation and Woollam\cite{Woollam71b}} & 0.41 & 40.3 & 98 & 0.41\\

\end{tabular}
\end{center}
\caption{Summary of fundamental frequencies and areas of the extremal orbits (in units of $10^{12}$ cm$^{-2}$) found
for the ellipsoidal and calculated SWM Fermi surface of graphite.}\label{tab1}
\end{table}

The minority carrier frequencies observed in graphite have been reviewed by Woollam~\cite{Woollam71b}. The area of the
neck orbits can easily be calculated at the $H$-point where the dispersion $E=\hbar v_f \sqrt{k_x^2+k_y^2}$ is linear.
The Fermi velocity $v_f=\sqrt{3} e a_0 \gamma_0/2\hbar$ depends only on $\gamma_0$, whose value of $3.15$~eV is
precisely known from magneto-optical data~\cite{Orlita2008,Chuang09,Ubrig2011}.  The area of $H$-point neck orbit for
$B \perp ab$ is $\pi k_f^2 = \pi E_f^2 / \hbar^2 v_f^2 = 0.43 \times 10^{12}$~cm$^{-2}$ corresponds to a frequency of
$\simeq 0.4$~T \emph{i.e.}, the large period ($1.2$~T$^{-1}$) minority carrier frequency of Ref.~\cite{Woollam71b}. The
neck orbits have their origin in the two interpenetrating hole ellipsoids at the H-point~\cite{Williamson65}. The size
of this orbit is expected to increase rapidly with tilt angle as the initially small circular orbit is transformed into
a large figure of eight orbit encompassing both hole ellipsoids (or a single ellipsoid if magnetic breakdown occurs at
the H-point)~\cite{Woollam71b}. The area of these neck orbits, with and without magnetic breakdown, have been
calculated within our simple model using the previously determined parameters for the majority hole ellipse. The only
adjustable parameter is the interpenetration of the hole ellipsoids which was chosen to have the correct minority
carrier frequency $\sim 0.4$~T. The calculated period of the neck orbits and are shown as dashed lines in
Fig.\ref{Fig2}. The neck orbits have the same frequency the spin split majority hole and spin split majority hole
harmonic at $\theta=90^{\circ}$ and so cannot be distinguished. Nevertheless, clear features correspond to the neck
orbit with magnetic breakdown are observed in the data for $75<\theta<90^{\circ}$.

Woollam assigned the minority carrier period of $\simeq 1$~T$^{-1}$ to the H-point neck orbits, which in view of our
results cannot be correct. Woollam's data is plotted as symbols in Fig.\ref{Fig2}(a) and (c) and seems to join up
nicely with the strong feature at around $50$~mT$^{-1}$ in our data. The calculated angular dependence for a neck orbit
and an ellipsoid are shown as broken lines in Fig.\ref{Fig2}(c). Clearly, the angular dependence corresponds to an
ellipsoid rather than a neck orbit. The angular dependence for an ellipsoid fitted to our data and the data of Woollam
is shown in Fig.\ref{Fig2}(a).

\begin{figure}
\begin{center}
\includegraphics[width= 4.5cm]{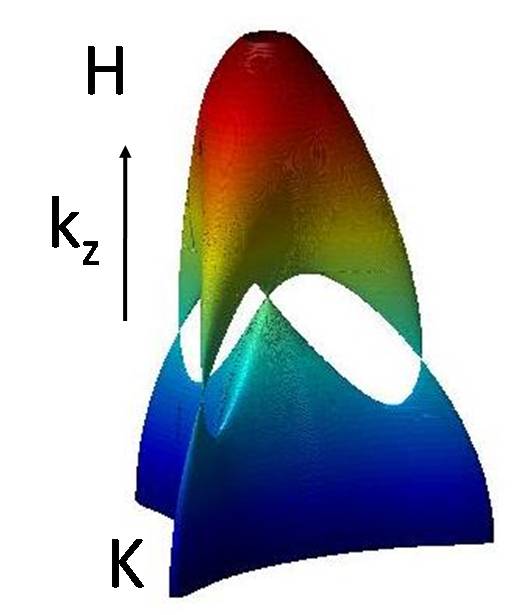}
\end{center}
\caption{(color online) SWM Fermi surface of graphite along the H-K-H edge.
}
 \label{Fig3}
\end{figure}

Finally, we have calculated the SWM Fermi surface. For the diagonalization the SWM matrix is truncated to a size of
$600 \times 600$. The magnetic field dependence of the density of states at $E_f$ is calculated at $k_z=0$ (electrons)
and $k_z=0.35$ (holes) assuming a reasonable Lorentzian broadening of the Landau levels. The Fourier transform is then
compared with the observed frequencies for $B\perp ab$. The SWM parameters $\gamma_0$ and $\gamma_1$ are precisely
known from magnetoptical data\cite{Orlita2008,Chuang09,Ubrig2011}. $E_f$ and $\gamma_3$ are treated as fitting
parameters. The hole surface is rather insensitive to $\gamma_3$ so that the correct hole frequency can be obtained by
choosing $E_f$, and then the electron frequency can be tuned using $\gamma_3$. After a few iterations this process
converges and the correct electron and hole frequencies are obtained with $E_f=-0.02505$~eV and $\gamma_3=0.443$~eV.
The SWM parameters used are summarized in Table \ref{tab2}. The Fermi surface is then calculated by diagonalizing the
SWM $4 \times 4$ matrix in zero magnetic field to calculate the in plane dispersion for $k_z=0-0.5$ and looking for the
crossing with $E_f$ for angles $\alpha=0-2\pi$ in the $k_x-k_y$ plane. The SWM Fermi surface is shown in Fig.\ref{Fig3}
and the calculated cross sectional areas are compared with the measured dHvA cross sections in Table \ref{tab1}. The
good agreement confirms that the diagonalisation of the truncated $600 \times 600$ matrix in magnetic field is fully
consistent with the results of diagonalizing the $4 \times 4$ SWM matrix in zero field.


While the calculated Fermi surface consistent with the majority electron and hole frequencies it cannot explain the
observed minority carrier frequency which is well approximated by an ellipsoid. Inspecting the SWM Fermi surface it can
be seen that there is no extremal orbit in the vicinity of $\theta=0$, so that the frequency should not be observed
except at high tilt angles, which is indeed the case for our data. However, this frequency is very clearly seen at
$\theta=0$ in the data of Woollam. This suggests that something is missing from the calculated Fermi surface so that a
significantly different set of SWM parameters might be required. Notably, increasing further the trigonal warping
parameter $\gamma_3$ can generate minority carrier pockets.

\begin{table}
\begin{center}
\begin{tabular}{ccc}
\hline\hline
$\gamma_0=3.15$~eV & $\gamma_1=0.375$~eV & $\gamma_2=-0.0243$~eV\\
$\gamma_3=0.443$~eV & $\gamma_4=0.07$~eV & $\gamma_5=0.05$~eV\\
$\Delta=-0.002$ & $E_f=-0.02505$~eV\\
\end{tabular}
\end{center}
\caption{Summary of the SWM parameters used.}\label{tab2}
\end{table}

In conclusion, angular dependent dHvA measurements on graphite reveal the 3D character of the Fermi surface of
graphite. The Fermi surfaces are closed in all directions and well approximated by elongated ellipsoids. Spin splitting
is clearly observed at high tilt angles and has to be included in the analysis in order to extract the correct Fermi
surface. The SWM parameter $\gamma_3$ is significantly larger than previously thought.

We would like to thank Yu. I. Latyshev for providing natural graphite samples.


\begin{thebibliography}{24}%
\makeatletter
\providecommand \@ifxundefined [1]{%
 \@ifx{#1\undefined}
}%
\providecommand \@ifnum [1]{%
 \ifnum #1\expandafter \@firstoftwo
 \else \expandafter \@secondoftwo
 \fi
}%
\providecommand \@ifx [1]{%
 \ifx #1\expandafter \@firstoftwo
 \else \expandafter \@secondoftwo
 \fi
}%
\providecommand \natexlab [1]{#1}%
\providecommand \enquote  [1]{``#1''}%
\providecommand \bibnamefont  [1]{#1}%
\providecommand \bibfnamefont [1]{#1}%
\providecommand \citenamefont [1]{#1}%
\providecommand \href@noop [0]{\@secondoftwo}%
\providecommand \href [0]{\begingroup \@sanitize@url \@href}%
\providecommand \@href[1]{\@@startlink{#1}\@@href}%
\providecommand \@@href[1]{\endgroup#1\@@endlink}%
\providecommand \@sanitize@url [0]{\catcode `\\12\catcode `\$12\catcode
  `\&12\catcode `\#12\catcode `\^12\catcode `\_12\catcode `\%12\relax}%
\providecommand \@@startlink[1]{}%
\providecommand \@@endlink[0]{}%
\providecommand \url  [0]{\begingroup\@sanitize@url \@url }%
\providecommand \@url [1]{\endgroup\@href {#1}{\urlprefix }}%
\providecommand \urlprefix  [0]{URL }%
\providecommand \Eprint [0]{\href }%
\providecommand \doibase [0]{http://dx.doi.org/}%
\providecommand \selectlanguage [0]{\@gobble}%
\providecommand \bibinfo  [0]{\@secondoftwo}%
\providecommand \bibfield  [0]{\@secondoftwo}%
\providecommand \translation [1]{[#1]}%
\providecommand \BibitemOpen [0]{}%
\providecommand \bibitemStop [0]{}%
\providecommand \bibitemNoStop [0]{.\EOS\space}%
\providecommand \EOS [0]{\spacefactor3000\relax}%
\providecommand \BibitemShut  [1]{\csname bibitem#1\endcsname}%
\let\auto@bib@innerbib\@empty
\bibitem [{\citenamefont {Slonczewski}\ and\ \citenamefont
  {Weiss}(1958)}]{Slonczewski58}%
  \BibitemOpen
  \bibfield  {author} {\bibinfo {author} {\bibfnamefont {J.~C.}\ \bibnamefont
  {Slonczewski}}\ and\ \bibinfo {author} {\bibfnamefont {P.~R.}\ \bibnamefont
  {Weiss}},\ }\href@noop {} {\bibfield  {journal} {\bibinfo  {journal} {Phys.
  Rev.}\ }\textbf {\bibinfo {volume} {109}},\ \bibinfo {pages} {272} (\bibinfo
  {year} {1958})}\BibitemShut {NoStop}%
\bibitem [{\citenamefont {McClure}(1960)}]{McClure60}%
  \BibitemOpen
  \bibfield  {author} {\bibinfo {author} {\bibfnamefont {J.~W.}\ \bibnamefont
  {McClure}},\ }\href@noop {} {\bibfield  {journal} {\bibinfo  {journal} {Phys.
  Rev.}\ }\textbf {\bibinfo {volume} {119}},\ \bibinfo {pages} {606} (\bibinfo
  {year} {1960})}\BibitemShut {NoStop}%
\bibitem [{\citenamefont {Nakao}(1976)}]{Nakao76}%
  \BibitemOpen
  \bibfield  {author} {\bibinfo {author} {\bibfnamefont {K.}~\bibnamefont
  {Nakao}},\ }\href@noop {} {\bibfield  {journal} {\bibinfo  {journal} {J.
  Phys. Soc. Japan}\ }\textbf {\bibinfo {volume} {40}},\ \bibinfo {pages} {761}
  (\bibinfo {year} {1976})}\BibitemShut {NoStop}%
\bibitem [{\citenamefont {Soule}(1958)}]{Soule58}%
  \BibitemOpen
  \bibfield  {author} {\bibinfo {author} {\bibfnamefont {D.~E.}\ \bibnamefont
  {Soule}},\ }\href@noop {} {\bibfield  {journal} {\bibinfo  {journal} {Phys.
  Rev.}\ }\textbf {\bibinfo {volume} {112}},\ \bibinfo {pages} {698} (\bibinfo
  {year} {1958})}\BibitemShut {NoStop}%
\bibitem [{\citenamefont {Soule}\ \emph {et~al.}(1964)\citenamefont {Soule},
  \citenamefont {McClure},\ and\ \citenamefont {Smith}}]{Soule64}%
  \BibitemOpen
  \bibfield  {author} {\bibinfo {author} {\bibfnamefont {D.~E.}\ \bibnamefont
  {Soule}}, \bibinfo {author} {\bibfnamefont {J.~W.}\ \bibnamefont {McClure}},
  \ and\ \bibinfo {author} {\bibfnamefont {L.~B.}\ \bibnamefont {Smith}},\
  }\href@noop {} {\bibfield  {journal} {\bibinfo  {journal} {Phys. Rev.}\
  }\textbf {\bibinfo {volume} {134}},\ \bibinfo {pages} {A453} (\bibinfo {year}
  {1964})}\BibitemShut {NoStop}%
\bibitem [{\citenamefont {Williamson}\ \emph {et~al.}(1965)\citenamefont
  {Williamson}, \citenamefont {Foner},\ and\ \citenamefont
  {Dresselhaus}}]{Williamson65}%
  \BibitemOpen
  \bibfield  {author} {\bibinfo {author} {\bibfnamefont {S.~J.}\ \bibnamefont
  {Williamson}}, \bibinfo {author} {\bibfnamefont {S.}~\bibnamefont {Foner}}, \
  and\ \bibinfo {author} {\bibfnamefont {M.~S.}\ \bibnamefont {Dresselhaus}},\
  }\href {\doibase 10.1103/PhysRev.140.A1429} {\bibfield  {journal} {\bibinfo
  {journal} {Phys. Rev.}\ }\textbf {\bibinfo {volume} {140}},\ \bibinfo {pages}
  {A1429} (\bibinfo {year} {1965})}\BibitemShut {NoStop}%
\bibitem [{\citenamefont {Schroeder}\ \emph {et~al.}(1968)\citenamefont
  {Schroeder}, \citenamefont {Dresselhaus},\ and\ \citenamefont
  {Javan}}]{Schroeder68}%
  \BibitemOpen
  \bibfield  {author} {\bibinfo {author} {\bibfnamefont {P.~R.}\ \bibnamefont
  {Schroeder}}, \bibinfo {author} {\bibfnamefont {M.~S.}\ \bibnamefont
  {Dresselhaus}}, \ and\ \bibinfo {author} {\bibfnamefont {A.}~\bibnamefont
  {Javan}},\ }\href@noop {} {\bibfield  {journal} {\bibinfo  {journal} {Phys.
  Rev. Lett.}\ }\textbf {\bibinfo {volume} {20}},\ \bibinfo {pages} {1292}
  (\bibinfo {year} {1968})}\BibitemShut {NoStop}%
\bibitem [{\citenamefont {Woollam}(1970)}]{Woollam70}%
  \BibitemOpen
  \bibfield  {author} {\bibinfo {author} {\bibfnamefont {J.~A.}\ \bibnamefont
  {Woollam}},\ }\href@noop {} {\bibfield  {journal} {\bibinfo  {journal} {Phys.
  Rev. Lett.}\ }\textbf {\bibinfo {volume} {70}},\ \bibinfo {pages} {811}
  (\bibinfo {year} {1970})}\BibitemShut {NoStop}%
\bibitem [{\citenamefont {Woollam}(1971{\natexlab{a}})}]{Woollam71a}%
  \BibitemOpen
  \bibfield  {author} {\bibinfo {author} {\bibfnamefont {J.~A.}\ \bibnamefont
  {Woollam}},\ }\href@noop {} {\bibfield  {journal} {\bibinfo  {journal} {Phys.
  Rev. B}\ }\textbf {\bibinfo {volume} {3}},\ \bibinfo {pages} {1148} (\bibinfo
  {year} {1971}{\natexlab{a}})}\BibitemShut {NoStop}%
\bibitem [{\citenamefont {Orlita}\ \emph {et~al.}(2008)\citenamefont {Orlita},
  \citenamefont {Faugeras}, \citenamefont {Martinez}, \citenamefont {Maude},
  \citenamefont {Sadowski},\ and\ \citenamefont {Potemski}}]{Orlita2008}%
  \BibitemOpen
  \bibfield  {author} {\bibinfo {author} {\bibfnamefont {M.}~\bibnamefont
  {Orlita}}, \bibinfo {author} {\bibfnamefont {C.}~\bibnamefont {Faugeras}},
  \bibinfo {author} {\bibfnamefont {G.}~\bibnamefont {Martinez}}, \bibinfo
  {author} {\bibfnamefont {D.~K.}\ \bibnamefont {Maude}}, \bibinfo {author}
  {\bibfnamefont {M.~L.}\ \bibnamefont {Sadowski}}, \ and\ \bibinfo {author}
  {\bibfnamefont {M.}~\bibnamefont {Potemski}},\ }\href {\doibase
  10.1103/PhysRevLett.100.136403} {\bibfield  {journal} {\bibinfo  {journal}
  {Phys. Rev. Lett.}\ }\textbf {\bibinfo {volume} {100}},\ \bibinfo {pages}
  {136403} (\bibinfo {year} {2008})}\BibitemShut {NoStop}%
\bibitem [{\citenamefont {Chuang}\ \emph {et~al.}(2009)\citenamefont {Chuang},
  \citenamefont {Baker},\ and\ \citenamefont {Nicholas}}]{Chuang09}%
  \BibitemOpen
  \bibfield  {author} {\bibinfo {author} {\bibfnamefont {K.-C.}\ \bibnamefont
  {Chuang}}, \bibinfo {author} {\bibfnamefont {A.~M.~R.}\ \bibnamefont
  {Baker}}, \ and\ \bibinfo {author} {\bibfnamefont {R.~J.}\ \bibnamefont
  {Nicholas}},\ }\href {\doibase 10.1103/PhysRevB.80.161410} {\bibfield
  {journal} {\bibinfo  {journal} {Phys. Rev. B}\ }\textbf {\bibinfo {volume}
  {80}},\ \bibinfo {pages} {161410} (\bibinfo {year} {2009})}\BibitemShut
  {NoStop}%
\bibitem [{\citenamefont {Schneider}\ \emph {et~al.}(2009)\citenamefont
  {Schneider}, \citenamefont {Orlita}, \citenamefont {Potemski},\ and\
  \citenamefont {Maude}}]{Schneider2009}%
  \BibitemOpen
  \bibfield  {author} {\bibinfo {author} {\bibfnamefont {J.~M.}\ \bibnamefont
  {Schneider}}, \bibinfo {author} {\bibfnamefont {M.}~\bibnamefont {Orlita}},
  \bibinfo {author} {\bibfnamefont {M.}~\bibnamefont {Potemski}}, \ and\
  \bibinfo {author} {\bibfnamefont {D.~K.}\ \bibnamefont {Maude}},\ }\href@noop
  {} {\bibfield  {journal} {\bibinfo  {journal} {Phys. Rev. Lett.}\ }\textbf
  {\bibinfo {volume} {102}},\ \bibinfo {pages} {166403} (\bibinfo {year}
  {2009})}\BibitemShut {NoStop}%
\bibitem [{\citenamefont {Zhu}\ \emph {et~al.}(2009)\citenamefont {Zhu},
  \citenamefont {Yang}, \citenamefont {Fauqu\'e}, \citenamefont {Kopelevich},\
  and\ \citenamefont {Behnia}}]{Zhu2009}%
  \BibitemOpen
  \bibfield  {author} {\bibinfo {author} {\bibfnamefont {Z.}~\bibnamefont
  {Zhu}}, \bibinfo {author} {\bibfnamefont {H.}~\bibnamefont {Yang}}, \bibinfo
  {author} {\bibfnamefont {B.}~\bibnamefont {Fauqu\'e}}, \bibinfo {author}
  {\bibfnamefont {Y.}~\bibnamefont {Kopelevich}}, \ and\ \bibinfo {author}
  {\bibfnamefont {K.}~\bibnamefont {Behnia}},\ }\href@noop {} {\bibfield
  {journal} {\bibinfo  {journal} {Nature Physics}\ }\textbf {\bibinfo {volume}
  {6}},\ \bibinfo {pages} {26} (\bibinfo {year} {2009})}\BibitemShut {NoStop}%
\bibitem [{\citenamefont {Schneider}\ \emph
  {et~al.}(2010{\natexlab{a}})\citenamefont {Schneider}, \citenamefont
  {Goncharuk}, \citenamefont {Vasek}, \citenamefont {Svoboda}, \citenamefont
  {Vyborny}, \citenamefont {Smrcka}, \citenamefont {Orlita}, \citenamefont
  {Potemski},\ and\ \citenamefont {Maude}}]{Schneider2010a}%
  \BibitemOpen
  \bibfield  {author} {\bibinfo {author} {\bibfnamefont {J.~M.}\ \bibnamefont
  {Schneider}}, \bibinfo {author} {\bibfnamefont {N.~A.}\ \bibnamefont
  {Goncharuk}}, \bibinfo {author} {\bibfnamefont {P.}~\bibnamefont {Vasek}},
  \bibinfo {author} {\bibfnamefont {P.}~\bibnamefont {Svoboda}}, \bibinfo
  {author} {\bibfnamefont {Z.}~\bibnamefont {Vyborny}}, \bibinfo {author}
  {\bibfnamefont {L.}~\bibnamefont {Smrcka}}, \bibinfo {author} {\bibfnamefont
  {M.}~\bibnamefont {Orlita}}, \bibinfo {author} {\bibfnamefont
  {M.}~\bibnamefont {Potemski}}, \ and\ \bibinfo {author} {\bibfnamefont
  {D.~K.}\ \bibnamefont {Maude}},\ }\href {\doibase 10.1103/PhysRevB.81.195204}
  {\bibfield  {journal} {\bibinfo  {journal} {Phys. Rev. B}\ }\textbf {\bibinfo
  {volume} {81}},\ \bibinfo {pages} {195204} (\bibinfo {year}
  {2010}{\natexlab{a}})}\BibitemShut {NoStop}%
\bibitem [{\citenamefont {Hubbard}\ \emph {et~al.}(2011)\citenamefont
  {Hubbard}, \citenamefont {Kershaw}, \citenamefont {Usher}, \citenamefont
  {Savchenko},\ and\ \citenamefont {Shytov}}]{Hubbard2011}%
  \BibitemOpen
  \bibfield  {author} {\bibinfo {author} {\bibfnamefont {S.~B.}\ \bibnamefont
  {Hubbard}}, \bibinfo {author} {\bibfnamefont {T.~J.}\ \bibnamefont
  {Kershaw}}, \bibinfo {author} {\bibfnamefont {A.}~\bibnamefont {Usher}},
  \bibinfo {author} {\bibfnamefont {A.~K.}\ \bibnamefont {Savchenko}}, \ and\
  \bibinfo {author} {\bibfnamefont {A.}~\bibnamefont {Shytov}},\ }\href
  {\doibase 10.1103/PhysRevB.83.035122} {\bibfield  {journal} {\bibinfo
  {journal} {Phys. Rev. B}\ }\textbf {\bibinfo {volume} {83}},\ \bibinfo
  {pages} {035122} (\bibinfo {year} {2011})}\BibitemShut {NoStop}%
\bibitem [{\citenamefont {Ubrig}\ \emph {et~al.}(2011)\citenamefont {Ubrig},
  \citenamefont {Plochocka}, \citenamefont {Kossacki}, \citenamefont {Orlita},
  \citenamefont {Maude}, \citenamefont {Portugall},\ and\ \citenamefont
  {Rikken}}]{Ubrig2011}%
  \BibitemOpen
  \bibfield  {author} {\bibinfo {author} {\bibfnamefont {N.}~\bibnamefont
  {Ubrig}}, \bibinfo {author} {\bibfnamefont {P.}~\bibnamefont {Plochocka}},
  \bibinfo {author} {\bibfnamefont {P.}~\bibnamefont {Kossacki}}, \bibinfo
  {author} {\bibfnamefont {M.}~\bibnamefont {Orlita}}, \bibinfo {author}
  {\bibfnamefont {D.~K.}\ \bibnamefont {Maude}}, \bibinfo {author}
  {\bibfnamefont {O.}~\bibnamefont {Portugall}}, \ and\ \bibinfo {author}
  {\bibfnamefont {G.~L. J.~A.}\ \bibnamefont {Rikken}},\ }\href {\doibase
  10.1103/PhysRevB.83.073401} {\bibfield  {journal} {\bibinfo  {journal} {Phys.
  Rev. B}\ }\textbf {\bibinfo {volume} {83}},\ \bibinfo {pages} {073401}
  (\bibinfo {year} {2011})}\BibitemShut {NoStop}%
\bibitem [{\citenamefont {Luk'yanchuk}\ and\ \citenamefont
  {Kopelevich}(2004)}]{Luk04}%
  \BibitemOpen
  \bibfield  {author} {\bibinfo {author} {\bibfnamefont {I.~A.}\ \bibnamefont
  {Luk'yanchuk}}\ and\ \bibinfo {author} {\bibfnamefont {Y.}~\bibnamefont
  {Kopelevich}},\ }\href@noop {} {\bibfield  {journal} {\bibinfo  {journal}
  {Phys. Rev. Lett.}\ }\textbf {\bibinfo {volume} {93}},\ \bibinfo {pages}
  {166402} (\bibinfo {year} {2004})}\BibitemShut {NoStop}%
\bibitem [{\citenamefont {Luk'yanchuk}\ and\ \citenamefont
  {Kopelevich}(2006)}]{Luk06}%
  \BibitemOpen
  \bibfield  {author} {\bibinfo {author} {\bibfnamefont {I.~A.}\ \bibnamefont
  {Luk'yanchuk}}\ and\ \bibinfo {author} {\bibfnamefont {Y.}~\bibnamefont
  {Kopelevich}},\ }\href@noop {} {\bibfield  {journal} {\bibinfo  {journal}
  {Phys. Rev. Lett.}\ }\textbf {\bibinfo {volume} {97}},\ \bibinfo {pages}
  {256801} (\bibinfo {year} {2006})}\BibitemShut {NoStop}%
\bibitem [{\citenamefont {Mikitik}\ and\ \citenamefont {{Yu. V.
  Sharlai}}(2006)}]{Mikitik06}%
  \BibitemOpen
  \bibfield  {author} {\bibinfo {author} {\bibfnamefont {G.~P.}\ \bibnamefont
  {Mikitik}}\ and\ \bibinfo {author} {\bibnamefont {{Yu. V. Sharlai}}},\
  }\href@noop {} {\bibfield  {journal} {\bibinfo  {journal} {Phys. Rev. B}\
  }\textbf {\bibinfo {volume} {73}},\ \bibinfo {pages} {235112} (\bibinfo
  {year} {2006})}\BibitemShut {NoStop}%
\bibitem [{\citenamefont {Luk'yanchuk}\ and\ \citenamefont
  {Kopelevich}(2010)}]{Lukyanchuk2010}%
  \BibitemOpen
  \bibfield  {author} {\bibinfo {author} {\bibfnamefont {I.~A.}\ \bibnamefont
  {Luk'yanchuk}}\ and\ \bibinfo {author} {\bibfnamefont {Y.}~\bibnamefont
  {Kopelevich}},\ }\href@noop {} {\bibfield  {journal} {\bibinfo  {journal}
  {Phys. Rev. Lett.}\ }\textbf {\bibinfo {volume} {104}},\ \bibinfo {pages}
  {119701} (\bibinfo {year} {2010})}\BibitemShut {NoStop}%
\bibitem [{\citenamefont {Schneider}\ \emph
  {et~al.}(2010{\natexlab{b}})\citenamefont {Schneider}, \citenamefont
  {Orlita}, \citenamefont {Potemski},\ and\ \citenamefont
  {Maude}}]{Schneider2010}%
  \BibitemOpen
  \bibfield  {author} {\bibinfo {author} {\bibfnamefont {J.~M.}\ \bibnamefont
  {Schneider}}, \bibinfo {author} {\bibfnamefont {M.}~\bibnamefont {Orlita}},
  \bibinfo {author} {\bibfnamefont {M.}~\bibnamefont {Potemski}}, \ and\
  \bibinfo {author} {\bibfnamefont {D.~K.}\ \bibnamefont {Maude}},\ }\href@noop
  {} {\bibfield  {journal} {\bibinfo  {journal} {Phys. Rev. Lett.}\ }\textbf
  {\bibinfo {volume} {104}},\ \bibinfo {pages} {119702} (\bibinfo {year}
  {2010}{\natexlab{b}})}\BibitemShut {NoStop}%
\bibitem [{\citenamefont {Nozi\`eres}(1958)}]{Nozieres58}%
  \BibitemOpen
  \bibfield  {author} {\bibinfo {author} {\bibfnamefont {P.}~\bibnamefont
  {Nozi\`eres}},\ }\href {\doibase 10.1103/PhysRev.109.1510} {\bibfield
  {journal} {\bibinfo  {journal} {Phys. Rev.}\ }\textbf {\bibinfo {volume}
  {109}},\ \bibinfo {pages} {1510} (\bibinfo {year} {1958})}\BibitemShut
  {NoStop}%
\bibitem [{\citenamefont {Woollam}(1971{\natexlab{b}})}]{Woollam71b}%
  \BibitemOpen
  \bibfield  {author} {\bibinfo {author} {\bibfnamefont {J.~A.}\ \bibnamefont
  {Woollam}},\ }\href@noop {} {\bibfield  {journal} {\bibinfo  {journal} {Phys.
  Rev. B}\ }\textbf {\bibinfo {volume} {4}},\ \bibinfo {pages} {3393} (\bibinfo
  {year} {1971}{\natexlab{b}})}\BibitemShut {NoStop}%
\bibitem [{\citenamefont {Onsager}(1952)}]{onsager1952}%
  \BibitemOpen
  \bibfield  {author} {\bibinfo {author} {\bibfnamefont {L.}~\bibnamefont
  {Onsager}},\ }\href@noop {} {\bibfield  {journal} {\bibinfo  {journal} {Phil.
  Mag. Ser. 7}\ }\textbf {\bibinfo {volume} {43}},\ \bibinfo {pages} {1006}
  (\bibinfo {year} {1952})}\BibitemShut {NoStop}%
\end{thebibliography}

%

\end{document}